\documentclass[twocolumn,preprintnumbers,aps]{revtex4}
\usepackage{graphicx}
\usepackage{makeidx}

\usepackage{color}

\newcommand{\be}{\begin{eqnarray}}
\newcommand{\ee}{\end{eqnarray}}

\begin{document}

\title{Condensation Energy of a Spacetime Condensate}

\author{Clovis Jacinto de Matos$^{\rm 1}$ Pavol Valko$^{\rm 2}$}

\affiliation{$^{\rm 1}$European Space Agency, 8-10 rue Mario
Nikis, 75015 Paris, France} \affiliation {$^{\rm 2}$ Dept. of
Physics, Slovak Technical University, Ilkovicova 3, 812 19
Bratislava, Slovakia}

\date{24 November 2010}

\preprint{}

\begin{abstract}
Starting from an analogy between the Planck-Einstein scale and the
dual length scales in Ginzburg-Landau theory of superconductivity,
and assuming that space-time is a condensate of neutral fermionic
particles with Planck mass, we derive the baryonic mass of the
universe. In that theoretical framework baryonic matter appears to
be associated with the condensation energy gained by spacetime in
the transition from its normal (symetric) to its (less symetric)
superconducting-like phase. It is shown however that the critical
transition temperature cannot be the Planck temperature. Thus
leaving open the enigma of the microscopic description of
spacetime at quantum level.
\end{abstract}

\maketitle

{\sc Introduction ---} The theory of superconductivity proposed by
Ginzburg and Landau in 1950 is based on Landau's theory of
second-order phase transition \cite{Tilley}. It was introduced as
a phenomenological theory, but later Gor'kov showed that it can be
derived from the full microscopic Bardeen Cooper and Schrieffer
(BCS) theory in a suitable limit. Since its introduction the
Ginzburg Landau theory was successfully used to describe
superfluidity in Fermi systems (${^3}He$) and other physical
systems where quantum phenomena are closely linked to phase
transition thermodynamics.

Spacetime undergoes several phase transition during its
cosmological evolution \cite{Sher}. Our assumption is that at
least one of them is a superconducting-like phase transition
driven by processes of spontaneous symmetry breaking. This type of
phase transition, allowed for any fermionic system exhibiting any
type of weak attractive interaction, would leave spacetime in a
condensed state with an energy gap and other critical parameters
like temperature marking the transition from the normal to the
superconducting state. Assuming also that gravitation is the
underlying bounding interaction of the spacetime condensate, then
further critical relevant parameters should be the condensate
density where higher density means higher critical temperature and
the gravitomagnetic field plays the role of the magnetic field in
superconductivity. To remain in the condensed state the difference
of free energy between the condensed and the normal spacetime
state should always remain negative and could be expressed in
terms of a thermodynamic critical gravitomagnetic field.

Having a spacetime condensate created in the process of a phase
transition it should be possible to apply the standard Ginzburg
Landau theory to spacetime. This means that two characteristic
lengths scales should exist for spacetime, which would be the
analog of the London penetration depth, and of the coherence
length. It is natural to assign the coherence length to Planck
length and the penetration length to the Einstein length
characterizing the size of the universe. This allows also to
associate a realistic spacetime energy density with the
thermodynamic energy density of the condensate, i.e. as the
geometric mean of two extreme energy densities, which we call the
Planck-Einstein scale. Doing this we calculate from this theory,
that the currently observed baryonic matter content of the
universe, corresponds to the spacetime condensation energy gained
in the transition from its normal to its new, less symetric state.

{\sc Condensation energy in superconductors} The great advantage
of the Ginzburg-Landau theory is its ability to solve many
difficult problems in superconductivity, without any reference to
the underlying microscopic BCS theory. In that sense Ginzburg
Landau theory is more general and could be used to describe other
solid state systems with spontaneously broken symmetry of ground
state. It could explain basic superconducting properties of exotic
superconductors, such as the high $T_c$ cuprates, even though the
original BCS theory does not seem to explain these systems.

We can analyze the phase diagram of superconductors in exactly the
same manner as one would consider the well known thermodynamics of
a liquid-gas phase transition problem such as given by the van der
Waals equations of state. However, for the superconductor instead
of the pair of thermodynamic variables $P$, $V$ (pressure and
volume) we have the magnetization $\vec M$ as the relevant
thermodynamic parameter.

The Gibbs free energy $G(T,H)$ is generally the most convenient
quantity to work with since the temperature $T$ and the external
magnetic field $H$ are the variables which are most naturally
controlled experimentally. Furthermore from $G(T,H)$ one can also
reconstruct the Helmoltz free energy $F(T,M)$, $F=G+\mu_0 V \vec H
. \vec M$, with $\mu_0$ being the magnetic permeability, and
volume $V$ the superconductor's volume. The difference in free
energies of superconducting and normal states at zero magnetic
 induction, $\vec M = -\vec H$, is
\begin{equation}
\Delta F_c=F_s(T,0)-F_n(T,0)=-\mu_0 V\frac{H_c^2}{2} \label{e1}
\end{equation}
The quantity $\mu_0 H_c^2/2$ is the condensation energy density.
It is a measure of the gain in free energy per unit volume in the
superconducting state compared with the normal state at the same
temperature $T$ below the critical transition temperature $T_c$.
The field $H_c^2$ is so-called thermodynamic critical field and
can be expressed as function of the coherence (healing) length
$\xi$, and of the London penetration depth $\lambda$.
\begin{equation}
H_c=\frac{\Phi_0}{2\pi \mu_0 \sqrt{2} \xi \lambda} \label{e2}
\end{equation}
where $\Phi_0=h/2e$ is the magnetic flux quantum ($h$ is the
Planck constant, and $e$ is the electron's electric charge).

Although the thermodynamic critical field isn't experimentally
observable it indicates how stable the condensed state is; and it
is also related with the observable critical fields, $H_{c1}$ and
$H_{c2}$ in Type-II superconductors, being their geometrical mean.

{\sc Planck-Einstein scale ---} The composition of the universe is
largely unknown to modern physics. Nucleosynthesis gives a rather
good estimate for the relative composition in baryonic matter. The
relative energy density of baryonic matter, $\Omega_B$ with
respect to the critical density,
\begin{equation}
\rho_c=3H_0^2 c^2/8\pi G\simeq 8.52\times10^{-10}
\quad[J/m^3]\label{cd}
\end{equation}
where $H_0\simeq2.3\times10^{-18} \quad [s^{-1}]$ is the Hubble
constant \cite{Hubble}, $G$ is universal gravitational constant,
and $c$ is the speed of light in vacuum; reveals that baryonic
matter only accounts for $4 \%$ of the total universe mass:
\begin{equation}
\Omega_B=\frac{\rho_B}{\rho_c}\sim0.04 \label{e4}
\end{equation}
Where $\rho_B$ is the density of observed baryonic matter in the
universe. The other $96\%$ of the universe mass is clearly unknown
and is allocated to the so called dark matter and dark energy.
Their respective relative densities with respect to the critical
energy density are respectively:
\begin{equation}
\Omega_{DM}=\frac{\rho_{DM}}{\rho_c}\sim0.23 \label{e5}
\end{equation}
and
\begin{equation}
\Omega_{DE}=\frac{\rho_{DE}}{\rho_c}\sim0.73 \label{e6}
\end{equation}

We are living in a universe that exhibits accelerating expansion
in (approximately) four space-time dimensions, with a de Sitter
spacetime metric with cosmological constant
$\Lambda=1.29\times10^{-52} [m^{-2}]$
\cite{Spergel,Peebles,Copeland,Padmanabhan,Padmanabhan2006}. A
small cosmological constant is equivalent to a small vacuum energy
density, which could account properly for the dark energy density
with equation of state $w=\rho/p=-1$ (showing a repulsive
interaction) given by
\begin{equation}
\rho_{vac \Lambda}=\frac{c^4\Lambda} {8 \pi
G}=6.21\times10^{-10}[J/m^3] \label{e7}.
\end{equation}
The fundamental scale, which could naturally host dark energy is
the \emph{Planck-Einstein scale}.

The Planck-Einstein scale corresponds to the geometric mean value
between the Planck scale, $l_P=(\hbar G/c^3)^{1/2}$, which
determines the highest possible energy density in the universe,
and the cosmological length scale, or Einstein scale,
$l_E=\Lambda^{-1/2}$, which determines the lowest possible energy
in the universe . The Planck scale is constructed out of the 4
fundamental constants, $c, \hbar, G, k$, where $\hbar=h/2\pi$ and
$k$ is Boltzmann constant. The Einstein scale is built from the
constants $c, \hbar, \Lambda, k$. Thus the Planck-Einstein length
$l_{PE}=\sqrt{l_P l_E}$ is the geometric mean of the two length
scales in the universe, and involves the five fundamental
constants $c, \hbar, G, \Lambda, k$. All other physical
Planck-Einstein scales for energy, mass, time, and density, are
calculated in a similar manner, Explicitly one has the following
formulas\cite{clovis}:

\begin{equation}
E_{PE}=kT_{PE}=\Bigg(\frac{c^7\hbar^3
\Lambda}{G}\Bigg)^{1/4}\label{e8}
\end{equation}
\begin{equation}
m_{PE}=\frac{E_{PE}}{c^2}=\Bigg(\frac{\hbar^3
\Lambda}{cG}\Bigg)^{1/4}\label{e9}
\end{equation}
\begin{equation}
l_{PE}=\frac{\hbar}{M_{PE}c}=\Bigg(\frac{\hbar G}{c^3
\Lambda}\Bigg)^{1/4}\label{e10}
\end{equation}
\begin{equation}
t_{PE}=\frac {l_{PE}}{c}=\Bigg(\frac{\hbar
G}{c^7\Lambda}\Bigg)^{1/4}\label{e11}
\end{equation}
\begin{equation}
\rho_{PE}=\frac{E_{PE}}{l_{PE}^3}=\frac{c^4 \Lambda}{G}\label{e12}
\end{equation}

One readily notices that the numerical values of Planck-Einstein
quantities correspond to typical time, length or energy scales in
superconductor physics, as well as to typical energy scales for
dark energy. In previous papers it has been pointed out
\cite{Beck3, Beck4, dematos} that there could be a deeper reason
for this coincidence: It is possible to construct theories of dark
energy that bear striking similarities with the physics of
superconductors. In these theories the Planck-Einstein scale
replaces the Planck scale as a suitable cutoff for vacuum
fluctuations.

{\sc Spacetime condensate and Baryonic Matter---} We will now draw
an analogy between the Ginzburg-Landau theory of
superconductivity, and the fundamental physical nature of
space-time at the Planck scale. Let us consider that space-time at
present time is an electrically neutral condensate created in a
phase transition coupled with spontaneous symmetry break-down
(possibly chiral symmetry breaking) in early period of universe
evolution. The two basic length scales of Ginzburg Landau theory
and space-time condensate should be associated as following: the
"London" penetration depth scale is equal to the Universe radius,
and the coherence length corresponds to the Planck length.
\begin{equation}
\lambda\equiv 1/\sqrt{\Lambda}\label{e13}
\end{equation}
\begin{equation}
\xi\equiv l_P \label{e14}
\end{equation}
Let us consider in addition that at the Planck scale, spacetime is
a condensate formed by pairs of neutral fermionic particles with
Planck mass $m_P=(\hbar c/G)^{1/2}$, which we call \emph{Planck's
condensate}. In that context the analog of the magnetic flux
quantum $\Phi_0$ is the gravitomagnetic flux quantum for Planck
mass $\Phi_{0g}$.
\begin{equation}
\Phi_{0g}=\frac{\hbar}{2m_P} \label{e15}
\end{equation}
The Planck condensate forms as soon as the universe cools down
below a certain transition temperature $\tau_c$, which we leave
undefined. In other words space-time becomes a condensate for
temperature  $T<\tau_c$. For temperatures higher than $\tau_c$
physical spacetime acquires full symmetry and ceases to be a
Planck's condensate.

The analog of the magnetic variable corresponding to the magnetic
field $\vec H$ for superconductors becomes the gravitomagnetic
field $\vec H_g$ in the case of a spacetime condensate (and when
gravitational field strength allows weak-field approximation). The
magnetization $\vec M$ is substituted by the gravitomagnetization
$\vec M_g$. The temperature $T$ remains thermodynamic variable in
both cases. The magnetic permeability $\mu_0$, must be accordingly
substituted by the gravitomagnetic permeability $\mu_{0g}=4\pi
G/c^2$ \cite{tajmar}. $V$ becomes the entire volume of the
observable universe $V_U$, which we assume to be spherical with
radius equal to the Einstein length $l_E=\Lambda^{-1/2}$.

The difference of free energy, $\Delta F_{gc}$, between
superconducting-like and normal states, at zero gravitomagnetic
field and zero gravitomagentization, $\vec M_g=-\vec H_g$, is:
\begin{equation}
\Delta F_{gc}=F_{gs}(T,0)-F_{gn}(T,0)=-\mu_{0g} V_U
\frac{H_{gc}^2}{4} \label{e17}
\end{equation}
This means that making the universe less symmetric (spacetime
condensate), will decrease its overall energy.

The quantity $\rho^*$
\begin{equation}
\rho^*=\mu_{0g} H_{gc}^2/4 \label{ee17}
\end{equation}
is the condensation energy density. It is a measure of the gain in
free energy per unit volume in the superconducting-like state
compared with the normal state at the same temperature $T$ below
the critical transition temperature $\tau_c$.

Substituting eq.(\ref{e13}), eq.(\ref{e14}), eq.(\ref{e15}), and
the gravitomagnetic permeability $\mu_{0g}$ in eq.(\ref{e2}), we
obtain the analog equation for the critical gravitomagnetic field
$H_{gc}$.
\begin{equation}
H_{gc}=\frac{1}{8\pi \sqrt{2}} \Big ( \frac{c^7 \Lambda}{G^3}
\Big)^{1/2}\frac{1}{m_P} \label{e18}
\end{equation}

Substituting the explicit expression of the Planck mass, in the
critical gravitomagnetic field, eq.(\ref{e18}), we obtain the
constant cosmological critical gravitomagnetic field expressed in
function of the universal fundamental constants, $G$, $\hbar$,
$c$, $\Lambda$.
\begin{equation}
H_{gc}=\frac{1}{8\pi \sqrt{2}} \Big( \frac {c^6\Lambda} {G^2} \Big
)^{1/2} \label{e20}
\end{equation}
Substituting eq.(\ref{e20}) into eq.(\ref{ee17}) we obtain the
constant cosmological density of condensation energy associated
with the emergence of the superconducting-like phase of spacetime:
\begin{equation}
\rho^*=\frac{1}{128 \pi}\frac{c^4\Lambda}{G}=3.88\times10^{-11}
\quad [J/m^3]\label{e21}
\end{equation}
One sees that $\rho^*$ is proportional to the Planck-Einstein
density of energy Eq.(\ref{e12}). Comparing the density of
condensation energy, eq.(\ref{e21}) with the critical density,
eq.(\ref{cd}) we find it to be in nearly exact coincidence with
the cosmological relative density of Baryonic matter,
eq.(\ref{e4}).
\begin{equation}
\Omega^*=\Omega_B=\frac{\rho^*}{\rho_c}=\frac{1}{48}\frac{c^2
\Lambda}{H_0^2}=0.0456\label{e22}
\end{equation}
Therefore if the condensation energy of spacetime is the Baryonic
matter content of our universe then the total excess of free
energy produced during the superconducting condensation of
space-time is the total mass of the Baryonic matter, $M_{BU}$,
present in our cosmos. Substituting eq.(\ref{e20}) in
eq.(\ref{e17}) and dividing by $c^2$ we obtain:
\begin{equation}
M_{BU}=\frac{\Delta F_{gc}}{c^2}=\frac{1}{96}\frac{c^2}{G
\Lambda^{1/2}}\sim1.24\times10^{51}\quad [Kg]\label{e23}
\end{equation}

{\sc Discussion and Conclusions ---} If we attempt to understand
the space-time condensation energy from the microscopic
composition of the Planck condensate we are directly lead into the
well known cosmological constant problem. This can be easily
demonstrated through an analogy with BCS theory: BCS theory shows
that the condensation energy density $\rho_F$ per atom in the
superconductor is of order:
\begin{equation}
\rho_F = kT_cg (\epsilon_F).\label{e24}
\end{equation}
Where $k$ is Boltzmann constant, $T_c$ is the critical temperature
and $g$ is the density of states at the Fermi level.

Since we considered in the previous section that spacetime
condensates when the universe cools down below the critical
temperature $\tau_c$, if we take $\tau_c$ equal to the Planck
temperature $T_P=k^{-1}(\hbar c^5/G)^{1/2}$, in eq.(\ref{e24}).
The analog of the condensation energy per atom, $\rho_F$ in
eq(\ref{e24}), will be played by the gained density of
condensation energy, i.e, the baryonic density of the universe
$\rho^*$, eq.(\ref{e21}).Carrying out this analogy eq.(\ref{e24})
becomes:
\begin{equation}
\rho^*=gE_P\label{e25}
\end{equation}
Multiplying both sides of eq.(\ref{e25}) by the Planck volume we
deduce that the baryonic energy density content of Planck cells is
125 orders of magnitude smaller than the total Planck energy
density:
\begin{equation}
\frac{\rho^* \l_P^3}{E_P}=g\l_p^3=8.37\times10^{-125}\label{e26}
\end{equation}
Since $\rho_{DE}\sim18 \rho^*$ it is easy to see that
eq.(\ref{e26}) is a possible statement of the so called
\emph{cosmological constant problem} \cite{bertolami}. This might
also indicate that the critical transition temperature $\tau_c$ of
the superconducting like spacetime seems not to be the Planck
temperature.

Presently we do not have yet a complete theory of quantum gravity
from which we could deduce the analog of density of states $g$ for
spacetime. Before BCS theory was formulated similar problems where
raised in the context of the Ginzburg Landau theory, to explain
the mysterious tiny condensation energy per atoms in
superconductors. Only BCS theory could explain that this energy is
so small because $kT_c$ is three orders of magnitude smaller than
the fermi energy, $\epsilon_F$. Will the theory of quantum gravity
be the equivalent of the BCS theory for superconductivity?
Although the use of Ginzburg-Landau concepts in the framework of
cosmology seems promising, this question is still open
\cite{Moffat}\cite{Stephon}.

\end{document}